\documentclass[lettersize,journal]{IEEEtran}
\usepackage{amsmath,amsfonts}
\usepackage{algorithmic}
\usepackage{algorithm}
\usepackage{array}
\usepackage[caption=false,font=normalsize,labelfont=sf,textfont=sf]{subfig}
\usepackage{textcomp}
\usepackage{stfloats}
\usepackage{url}
\usepackage{verbatim}
\usepackage{graphicx}
\usepackage{cite}
\usepackage{booktabs}
\usepackage{multirow} 
\usepackage{enumitem}
\usepackage{color}
\usepackage[colorlinks,
            linkcolor=blue,
            anchorcolor=blue,
            citecolor=blue]{hyperref}

\begin{document}

\title{RA-BLIP: Multimodal Adaptive Retrieval-Augmented Bootstrapping Language-Image Pre-training}

\author{Muhe Ding, Yang Ma, Pengda Qin, Jianlong Wu,~\IEEEmembership{Member,~IEEE,} Yuhong Li, Liqiang Nie,~\IEEEmembership{Senior Member,~IEEE}

\thanks{This work was supported in part by the National Natural Science Foundation of China under Grant 62376069, in part by Young Elite Scientists Sponsorship Program by CAST under Grant 2023QNRC001, and in part by Guangdong Basic and Applied Basic Research Foundation under Grant 2024A1515012027.}
    \thanks{
    Muhe Ding, Jianlong Wu and Liqiang Nie are with the School of Computer Science and Technology, Harbin Institute of Technology (Shenzhen), Shenzhen 518055, China (e-mail: dmh1216380870@gmail.com, jlwu1992@pku.edu.cn, nieliqiang@gmail.com).
    }
    \thanks{Yang Ma is with the School of Computer Science, University of Sydney, Sydney, NSW 2006, Australia (e-mail: yama5878@uni.sydney.edu.au).}
    \thanks{Pengda Qin and Yuhong Li are with the Security Department, Alibaba Group, Hangzhou 311121, China (e-mail: qinpengda0406@163.com, daniel.yuhong@gmail.com).}
    
    }

\markboth{IEEE TRANSACTIONS ON MULTIMEDIA}%
{Shell \MakeLowercase{\textit{et al.}}: A Sample Article Using IEEEtran.cls for IEEE Journals}


\maketitle

\begin{abstract}

Multimodal Large Language Models~(MLLMs) have recently received substantial interest, which shows their emerging potential as general-purpose models for various vision-language tasks. 
MLLMs involve significant external knowledge within their parameters; however, it is challenging to continually update these models with the latest knowledge, which involves huge computational costs and poor interpretability. Retrieval augmentation techniques have proven to be effective plugins for both LLMs and MLLMs. In this study, we propose multimodal adaptive Retrieval-Augmented Bootstrapping Language-Image Pre-training~(\textbf{RA-BLIP}), a novel retrieval-augmented framework for various MLLMs. Considering the redundant information within vision modality, we first leverage the question to instruct the extraction of visual information through interactions with one set of learnable queries, minimizing irrelevant interference during retrieval and generation. 
Besides, we introduce a pre-trained multimodal adaptive fusion module to achieve question text-to-multimodal retrieval and integration of multimodal knowledge by projecting visual and language modalities into a unified semantic space. Furthermore, we present an Adaptive Selection Knowledge Generation~(ASKG) strategy to train the generator to autonomously discern the relevance of retrieved knowledge, which realizes excellent denoising performance. Extensive experiments on open multimodal question-answering datasets demonstrate that RA-BLIP achieves significant performance and surpasses the state-of-the-art retrieval-augmented models.
\end{abstract}


\section{Introduction}
\label{sec:intro}

\begin{figure}[t]
  \centering
   \includegraphics[width=0.9\linewidth]{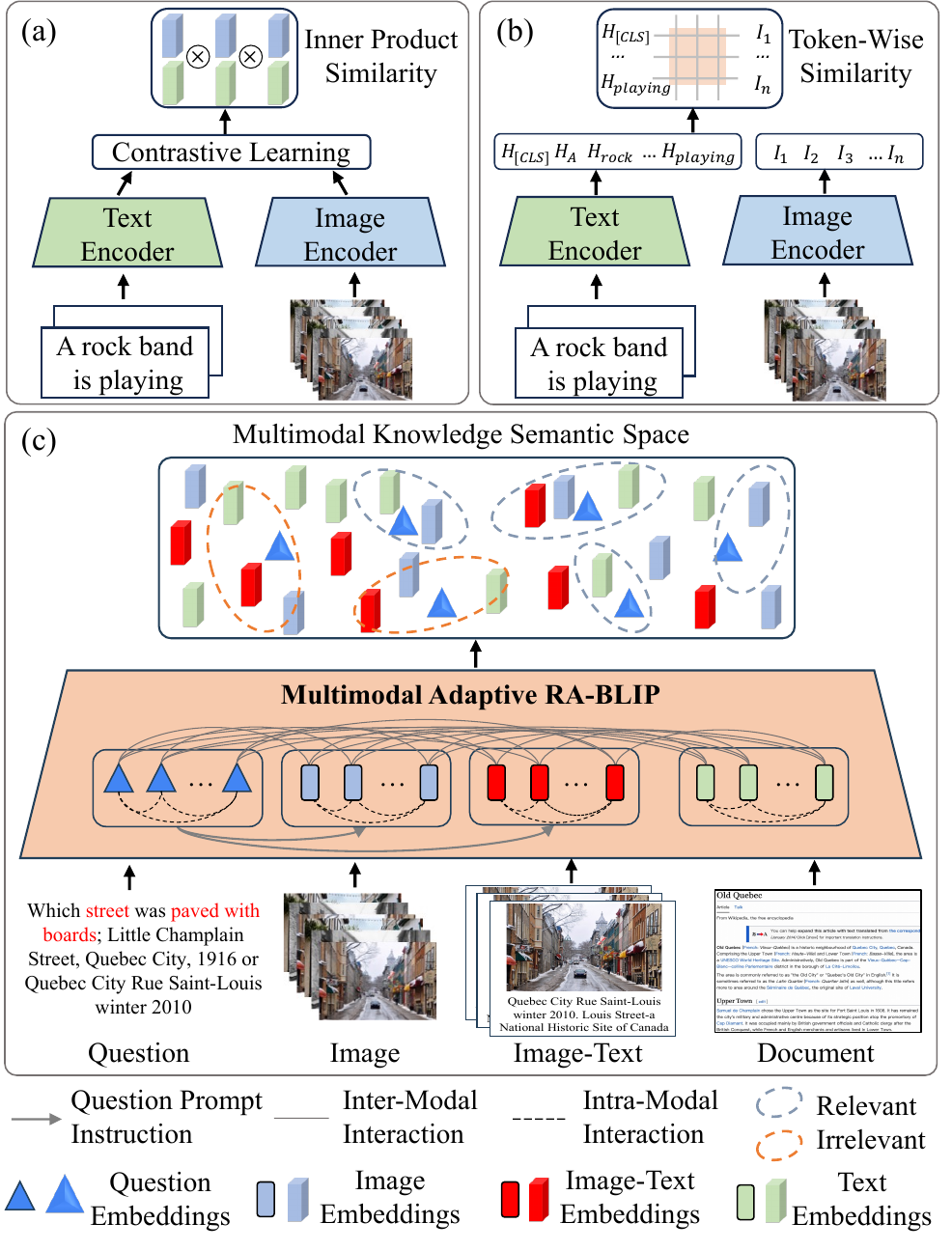}
   \caption{Illustration of different multimodal retrieval approaches. (a) Cross-modality retrieval. (b) Late-interaction retrieval. (c) RA-BLIP multimodal adaptive retrieval. For RA-BLIP, questions, documents, images, and image-text pairs are projected into a unified multimodal space.}
   \label{Introduction}
\end{figure}

\IEEEPARstart{T}{he} birth of the Internet has triggered an unprecedented information revolution, catapulting humanity into the era of information explosion. It is a great challenge to efficiently find answers from a vast amount of information based on our questions. Open Multimodal Multihop Question Answering~(MMQA)~\cite{DBLP:conf/iccv/AntolALMBZP15,DBLP:conf/cvpr/GoyalKSBP17,DBLP:conf/acl/LeeCT19,DBLP:conf/cvpr/ChangCNGSB22,DBLP:conf/iclr/TalmorYCLWAIHB21,DBLP:conf/acl/LiLN22,DBLP:journals/tmm/WuZQ24} can help alleviate this problem of information overload by retrieving external knowledge based on questions and generating correct answers. In recent years, several advanced LLMs and MLLMs like FlanT5~\cite{DBLP:journals/corr/abs-2210-11416}, LLaMA~\cite{DBLP:journals/corr/abs-2302-13971}, BLIP-2~\cite{DBLP:conf/icml/0008LSH23}, GPT-4~\cite{DBLP:journals/corr/abs-2303-08774}, etc., have been notably explored to enhance their performance by implicitly encoding a substantial amount of external knowledge within their parameters, which now scale into the hundreds of billions~\cite{DBLP:journals/jmlr/ChowdheryNDBMRBCSGSSTMRBTSPRDHPBAI23}. While these models have yielded exciting results on various multimodal tasks, they have also encountered high computational costs and significant challenges in terms of interpretability. 

To alleviate the challenge, many researchers proposed retrieval augmentation techniques that divide the model into two key components: the retriever and the generator~\cite{DBLP:conf/icml/GuuLTPC20,DBLP:conf/emnlp/ChenHCVC22,DBLP:conf/cvpr/HuI0WCSSRF23,DBLP:conf/mm/RaoSLZY23}. The retriever accesses relevant knowledge from a knowledge base based on the posed question, while the generator leverages this information to create textual output in response. In earlier stages, text-modality retrieval-augmented models, such as  REALM~\cite{DBLP:conf/icml/GuuLTPC20}, RAG~\cite{DBLP:conf/nips/LewisPPPKGKLYR020}, and so on~\cite{DBLP:conf/eacl/IzacardG21,DBLP:conf/icml/BorgeaudMHCRM0L22}, have been proposed to solve text-only question answering. They build the dense index as a non-parametric document memory from extensive textual sources like Wikipedia for effective knowledge retrieval, and the generator produces answers based on the retrieved knowledge. More recently, multimodal retrieval-augmented models, such as MuRAG~\cite{DBLP:conf/emnlp/ChenHCVC22}, SKURG~\cite{DBLP:journals/corr/abs-2212-08632}, and so on~\cite{DBLP:conf/acl/0002FYHL23,DBLP:conf/mm/RaoSLZY23}, have emerged one after another. These models extend the knowledge memory across various modalities, employing pre-trained visual language models to retrieve relevant evidence and support reasoning for answers.  

However, existing methods exhibit certain limitations. The first limitation is the insufficient integration and interaction between vision and language. On the one hand, existing methods lack explicit integration of multimodal information, hindering the alignment of questions and multimodal knowledge in the semantic space. As shown in Fig.~\ref{Introduction}(a), some methods~\cite{DBLP:journals/tmm/LiuLFHL21,DBLP:conf/emnlp/ChenHCVC22,DBLP:journals/corr/abs-2209-00179} have employed separate visual encoder and text encoder for individual modality encoding and adopted contrastive learning~\cite{DBLP:conf/icml/ChenK0H20} for multimodal alignment to retrieve. This may lead to an unbalanced and biased multimodal retrieval and reasoning process towards specific modalities. Besides, late-interaction retrieval approaches~\cite{DBLP:conf/acl/LuZL20,DBLP:conf/emnlp/LiuYL21,DBLP:conf/sigir/LinMLSWN23} in Fig.~\ref{Introduction}(b), retain dual-encoder independent encoding architecture and perform token-wise interactions only in the late scoring stage, which sacrifices the retrieval efficiency for the benefits of fine-grained feature learning. On the other hand, existing methods~\cite{DBLP:journals/tmm/YuZLQHTW20,DBLP:conf/emnlp/ChenHCVC22,DBLP:journals/corr/abs-2212-08632,DBLP:journals/tmm/ZhuangYDQH24} do not utilize questions to instruct the image encoder in selectively extracting visual features, and thus suffer from interference and noise caused by redundant information in images. Moreover, such methods lack mutual instruction when encoding different modal features, making it challenging to model relationships between multiple sources. The second limitation is that existing approaches do not inspect the correctness of the retrieved relevant knowledge at the generation stage. However, the retrieved knowledge contains significant noise, resulting in poor model anti-interference and robustness. The generator assumes all retrieved relevant knowledge is correct, potentially leading to the utilization of irrelevant or confusing information.

To address the above issues, we propose multimodal adaptive Retrieval-Augmented Bootstrapping Language-Image Pre-training~(RA-BLIP). RA-BLIP consists of two key components: a multimodal adaptive retrieval-augmented framework and an adaptive selection knowledge generation~(ASKG) strategy. To tackle the first limitation, RA-BLIP is based on the InstructBLIP architecture~\cite{DBLP:journals/corr/abs-2305-06500} and adopts Q-Former to implement instruction-aware visual feature extraction that uses questions as instructions. The question instruction interacts with the query embeddings through shared self-attention layers and encourages the extraction of question-relevant visual features. Additionally, we incorporate a pre-trained multimodal adaptive fusion module to fuse vision and text information, obtaining multimodal features~\cite{DBLP:conf/nips/VaswaniSPUJGKP17,DBLP:conf/naacl/DevlinCLT19,DBLP:conf/www/ZhengWLXZLZ23}. As a result, RA-BLIP achieves question text-to-multimodal retrieval by aligning the questions and multimodal knowledge bases in the semantic space of three modalities: text, image, and image-text~\cite{DBLP:conf/kdd/YuCSWCBZ22}, as shown in Fig.~\ref{Introduction}(c). For the second limitation, we leverage the implicit capabilities of LLMs and introduce an adaptive selection knowledge generation strategy, which gives the generator the capability of selecting knowledge by data enhancement to make the model automatically judge the relevance of knowledge. ASKG strategy allows the generator to not simply rely on the word similarity between the question and knowledge, but to understand the semantic information of question and know which knowledge contains the answer. Furthermore, the parameters of the image encoder and LLM of our framework are frozen, significantly reducing computational costs. Extensive experiments on three representative QA datasets demonstrate the effectiveness of our methods.

Overall, our key contributions are as follows:

\begin{itemize}
    \item{We propose a novel multimodal adaptive retrieval-augmented framework, which achieves question text-to-multimodal retrieval and knowledge-intensive multimodal QA by integrating visual and language modalities and projecting them into a unified semantic space.}

    \item{We introduce an adaptive selection knowledge generation strategy that leverages the powerful capabilities of LLMs to select the relevant retrieved knowledge for answer reasoning autonomously.}

    \item{We conduct extensive experiments on various multimodal and multihop datasets (i.e., WebQA~\cite{DBLP:conf/cvpr/ChangCNGSB22}, MultimodalQA~\cite{DBLP:conf/iclr/TalmorYCLWAIHB21}, and MMCoQA~\cite{DBLP:conf/acl/LiLN22}). RA-BLIP demonstrates superiority over the existing state-of-the-art retrieval-augmented models.}
    
\end{itemize}

\section{Related Work}
\label{sec:relatedwork}

\subsection{Vision-Language Pretraining}
Vision-language pre-training (VLP) aims to train models on large-scale image-text datasets to capture the relationship between these two modalities. Broadly, VLP methodologies fall into two categories based on their training approach:
1) End-to-end Methods:  This category includes methods \cite{ DBLP:conf/icml/JiaYXCPPLSLD21, DBLP:conf/emnlp/TanB19, DBLP:conf/iclr/WangYYDT022,DBLP:journals/tmm/QiZLSWLL23} that train models end-to-end, backpropagating learned signals to achieve mutual learning between different modalities.
2) Modular Methods: In contrast, modular methods, as seen in works by \cite{DBLP:conf/eccv/ChenLYK0G0020, DBLP:conf/eccv/Li0LZHZWH0WCG20, DBLP:journals/corr/abs-2101-00529, DBLP:conf/nips/TsimpoukelliMCE21,DBLP:journals/tmm/XingWCZLWZ24}, involve keeping the parameters of specific pre-trained components (like image encoders or large language models) fixed while focusing on refining other aspects of the model.
For instance, LiT \cite{DBLP:conf/cvpr/ZhaiWMSK0B22} utilizes a pre-trained frozen image encoder from CLIP, while Flamingo \cite{DBLP:conf/nips/AlayracDLMBHLMM22} and BLIP-2 \cite{DBLP:conf/icml/0008LSH23} freeze the language model to integrate LLMs into vision-language tasks better.
Besides, instruction tuning is also an effective approach during VLP. InstructBLIP \cite{DBLP:journals/corr/abs-2305-06500} represents a recent advancement in this area, achieving instruction-aware visual feature extraction and instruction-guided LLM generation.
The unique capability of InstructBLIP to extract features based on prompt instructions, combined with its utilization of frozen LLMs and image encoders, positions it as an ideal backbone for our proposed retrieval-augmented framework.

\subsection{Text-modality Retrieval-Augmented Models}
Retrieval-augmented techniques have proven to be effective plugins for both LLMs and MLLMs in academia. 
These techniques extract pertinent world knowledge from extensive databases, subsequently integrating this information to formulate answers.
Pioneering methods like ORQA \cite{DBLP:conf/acl/LeeCT19} have employed inverse cloze tasks for retriever pre-training, showcasing their efficacy on open-ended question answering datasets.
Following suit, REALM \cite{DBLP:conf/icml/GuuLTPC20} extends this by retrieving and processing documents from comprehensive sources like Wikipedia for logical reasoning, which is pre-trained to reason over a large corpus of
knowledge on the fly during inference. Furthermore, RAG \cite{DBLP:conf/nips/LewisPPPKGKLYR020} adopts a pre-trained model and non-parametric memory for language generation.
FiD \cite{DBLP:conf/eacl/IzacardG21} leverages encoder-decoder transformer models for knowledge-intensive tasks, setting new benchmarks in QA. More recently, RETRO \cite{DBLP:conf/icml/BorgeaudMHCRM0L22} has advanced these methods by handling longer sequences and accessing diverse documents for segmented sequences from expansive retrieval datasets. Text-modality retrieval-augmented models construct dense indices as non-parametric document memories, using extensive textual knowledge bases to align the retrieval process with specific queries. Despite these advancements, a notable limitation remains the need for these methods to effectively leverage vast multimodal knowledge, thereby constraining their applicability in the domain of open multimodal question answering.

\subsection{Multimodal Retrieval-Augmented Models}
To overcome the limitations of text-modality retrieval-augmented models, recent research \cite{DBLP:conf/emnlp/0003LSM21, DBLP:conf/acl/YavuzHZKX22, DBLP:conf/cvpr/HuI0WCSSRF23, DBLP:conf/icml/0008LSH23, DBLP:journals/corr/abs-2305-06500} has made strides in integrating multimodal knowledge. 
Notable efforts, including AutoRouting \cite{DBLP:conf/iclr/TalmorYCLWAIHB21} and MAE \cite{DBLP:conf/acl/LiLN22}, involve training distinct models for each modality and using classifiers for task-specific routing, though this approach often hampers cross-modal reasoning. MuRAG \cite{DBLP:conf/emnlp/ChenHCVC22} seeks to overcome this limitation by employing separate encoders for visual and textual modalities, followed by a joint encoder for multimodal fusion. However, this approach lacks integrated guidance for different modalities and cannot model the relations between knowledge sources during retrieval. SKURG~\cite{DBLP:journals/corr/abs-2212-08632} attempts to bridge this gap by using an entity-centered fusion encoder to align modalities, yet faces challenges in computational efficiency and limited interpretability. Besides, Solar~\cite{DBLP:conf/acl/0002FYHL23} transforms multimodal inputs into a unified language format but falls short in handling complex tasks and generalizing visual information.
REVAL~\cite{DBLP:conf/mm/RaoSLZY23} leverages large-scale knowledge graphs to assist visual language pre-training, but it brings a lot of calculations.
In contrast, RA-BLIP distinguishes itself by seamlessly integrating visual and language modalities into a cohesive semantic space, enabling the autonomous selection of relevant knowledge for reasoning, thus addressing these limitations more effectively.


\section{METHODOLOGY}

In this section, we first formulate the research problem and subsequently elaborate on the model architecture of our retrieval-augmented framework. Then, we describe the learnable query interaction approach, followed by multimodal adaptive fusion module. Subsequently, we show how to train the retriever and rank the relevant knowledge. Lastly, we introduce the adaptive selection knowledge generation strategy.

\begin{figure*}[!t]
  \centering
   \centerline{\includegraphics[width=1\linewidth]{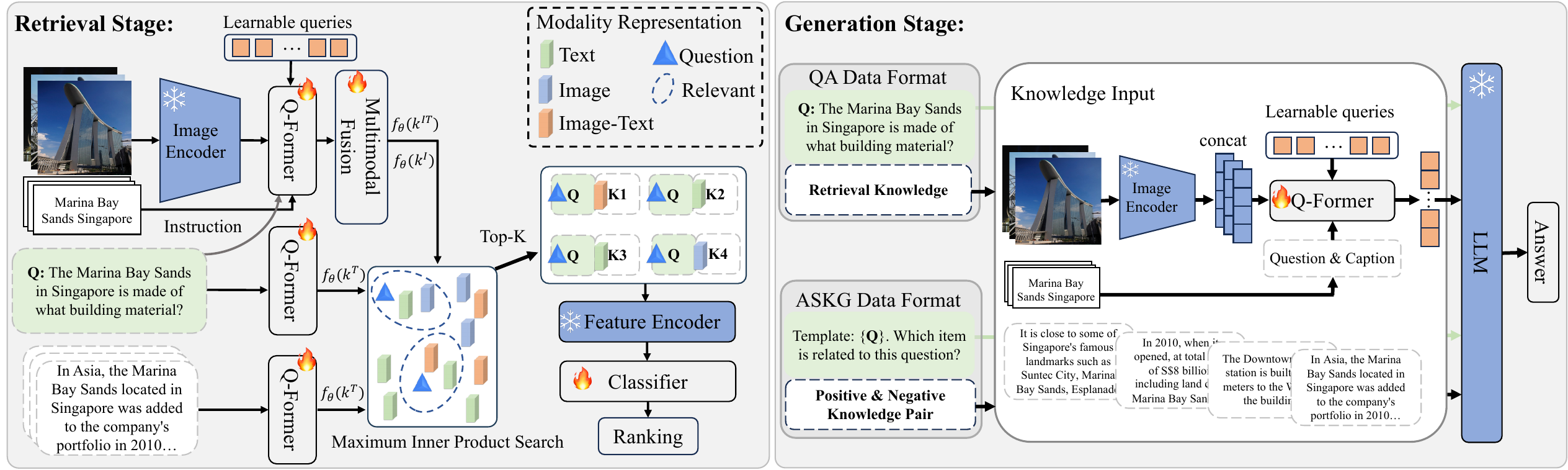}}
   \caption{The overall workflow of RA-BLIP consists of a retrieval stage and a generation stage. We utilize multimodal encoder $f_{\theta}(\cdot)$ to project questions and multimodal knowledge into a unified semantic space to achieve question text-to-multimodal retrieval, and exclude confusing knowledge by ranking.
   Additionally, we employ ASKG strategy to filter out invalid knowledge, enabling precise reasoning. The parameters of LLM and image encoder are fixed, only the Q-Former and classifier are trainable.}
   \label{The framework}
\end{figure*}

\subsection{Problem Formulation}\label{3.1}
This paper presents a multimodal adaptive retrieval-augmented framework called RA-BLIP for open multihop and multimodal QA, integrating retrieval and generation functions. For knowledge-intensive QA, we deconstruct the task into two stages: retrieval and generation, which are implemented by the retriever and generator respectively. The goal of our model training is to learn the distribution $P(y|x_q)$ to generate a textual output $y$ conditioned on input question $x_q$ and multimodal knowledge base $\mathcal{KB}$~($\mathcal{KB} = {k_1,...,k_n}$). Firstly, the retriever encodes questions, images, and texts from the knowledge base $\mathcal{KB}$. It identifies the most relevant retrieved knowledge, $\mathcal{K}_{ret} \subset \mathcal{KB}$~($\mathcal{K}_{ret}$ is the retrieved knowledge) for each question $x_q$, which is modeled as $p(\mathcal{K}_{ret}| x_q)$. Secondly, the generator utilizes an LLM to generate answers $y$, conditioned on both the question and the retrieved knowledge, which is modeled as $p(y|x_q,\mathcal{K}_{ret})$. We treat multimodal knowledge $\mathcal{K}_{ret}$ as a latent variable from the external knowledge base and marginalize it to increase the overall likelihood of the answer $y$. The overall process is encapsulated in the equation: 
\begin{equation}
  p(y \mid x_q)=\sum_{\mathcal{K}_{ret} \subset \mathcal{KB}} \underbrace{p(\mathcal{K}_{ret} \mid x_q)}_{\text {Retrieval}} \cdot \underbrace{p(y \mid x_q, \mathcal{K}_{ret})}_{\text {Generation}}.
  \label{eq1}
\end{equation}
This dual-stage framework effectively addresses the complexities of open multimodal QA by balancing the retrieval of multimodal data and knowledge-based generation, and has been validated by extensive experiments and ablation studies.

\subsection{Model Architecture}\label{3.2}

RA-BLIP is built on a simple backbone model that is pre-trained to encode image-text pairs so that they are suitable for both knowledge base retrieval and answer generation. The overall framework of RA-BLIP is shown in Fig.~\ref{The framework}.
The backbone model consists of a multimodal encoder $f_{\theta}(\cdot)$ and decoder $g_{\theta}(\cdot)$, which are used as components of the RA-BLIP model to implement retrieval and generation. The multimodal encoder $f_{\theta}(\cdot)$ contains a frozen image encoder ViT~\cite{DBLP:conf/iclr/DosovitskiyB0WZ21}, Q-Former architecture~\cite{DBLP:journals/corr/abs-2305-06500}, and the pre-trained multimodal adaptive fusion module. 
The decoder $g_{\theta}(\cdot)$ is composed of a LLM FlanT5~\cite{DBLP:journals/corr/abs-2210-11416}.
Querying Transformer~(Q-Former)~\cite{DBLP:conf/icml/0008LSH23} is a lightweight Transformer consisting of two modules that share the same self-attention layer: one is an image transformer that interacts with the frozen image encoder ViT for visual feature extraction, and the other is a text transformer can act as both text encoder and text decoder for text feature. The visual encoder, composed of ViT and Q-Former image transformer, has instruction-aware visual feature extraction capabilities and can extract visual information based on question instructions. We input $N$ learnable query embeddings into the Q-Former image transformer, which interacts with frozen image features through cross-attention layers to obtain visual representation $f_{\theta}(I)\in\mathbb{R}^{N \times D}$, where $D$ is the hidden dimension of the Q-Former. Additionally, we use the Q-Former text transformer to encode text, taking the $\operatorname{[CLS]}$ token as the text representation $f_{\theta}(T) \in \mathbb{R}^{1 \times D}$.  To obtain multimodal features combining both image and text, we introduce a pre-trained multimodal adaptive fusion module $M_\theta(\cdot)$ to obtain the multimodal representation $f_{\theta}(IT)\in\mathbb{R}^{(N+1) \times D}$. Consequently, the multimodal encoder can simultaneously encode image, text, and image-text features.  Questions and knowledge are encoded into multimodal information through the multimodal encoder and then input into the decoder for answer generation. In the generation stage, compared with the retriever, the multimodal encoder discards the multimodal adaptive fusion module to reduce the computational cost.

\subsection{Learnable Query Interaction for Multi-images}\label{3.3}

Questions and image captions are used as instructions to extract visual features to get learnable query embeddings and input them together with text knowledge to LLMs for generation. We employ a novel approach for extracting visual information to alleviate the burden of LLMs in distinguishing knowledge.
In the original Q-Former in~\cite{DBLP:journals/corr/abs-2305-06500}, multiple images are processed by employing multiple separate sets of queries for each, with each set of queries independently extracting visual features. This results in using multiple query features for generation, which can be computationally intensive and less efficient in capturing the interrelations among different images. In contrast, our method innovates by succinctly utilizing one set of learnable queries to directly interact with and extract features from multiple images in a unified manner. This process occurs during the Q-Former stage, enabling more efficient and integrated interaction among multiple visual references. 
By employing one set of query interaction approach, RA-BLIP not only simplifies the feature extraction process but also enhances the efficiency of information extraction. This unified interaction allows the model to understand better and represent the collective information presented in multiple images, enabling more effective and cohesive feature utilization, especially when dealing with complex scenes or subjects across multiple images. 


\subsection{Multimodal Adaptive Fusion Module}\label{MAFM}

\begin{figure}[t]
  \centering
   \includegraphics[width=1\linewidth]{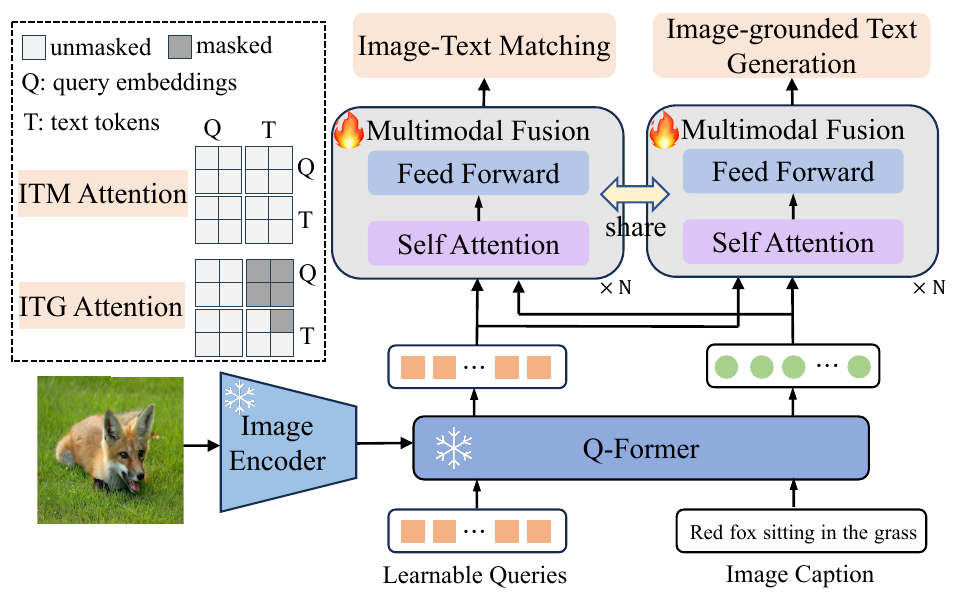}
   \caption{Schematic diagram of the pre-training process of multimodal adaptive fusion module.}
   \label{Frozen}
\end{figure}

The pre-trained multimodal adaptive fusion module $M_\theta(\cdot)$ consists of a 3-layer BERT network~\cite{DBLP:conf/nips/VaswaniSPUJGKP17,DBLP:conf/naacl/DevlinCLT19}. Since the Q-Former has aligned visual and text feature representation, we use Image-Text Matching loss and Image-grounded Text Generation loss for pre-training. As shown in Fig.~\ref{Frozen}, our approach is to fix the parameters of the image encoder and Q-Former, and solely fine-tune the parameters of the multimodal adaptive fusion module. The module concatenates the visual embedding and text embedding with a dimension of $\mathbb{R}^{(N+L)\times D}$, where $N$ is the learnable
query embeddings and $L$ is the length of text tokens. Image-text matching loss is used to fuse image and text representations, and the results of the fusion module are fed into a binary linear classifier for each output query to obtain the logits and take the average logits of all queries as the matching score. Given a pre-training dataset $\mathcal{X}=\{I_i,T_i\}^n_{i=1}$, we randomly sample negative texts for each image and randomly sample negative images for each text, to generate negative training data. Therefore, we denote the ground truth label as $y \in \{1, 0\}$ for each image-text pair $(I_i,T_i)$, indicating if the input image-text pair is relevant or not. We use the multimodal encoder $f_\theta(\cdot)$ to encode image-text pairs and input it into the multimodal adaptive fusion module $M_\theta(\cdot)$.
The objective function is defined as follows:
\begin{equation}
\mathcal{L}_{itm}=-\frac{1}{n}\sum_{I_i, T_i \in \mathcal{X}} y \log \left(\rho\left(M_\theta\left(f_\theta(I_i) ; f_\theta(T_i)\right)\right)\right),
\end{equation}
where $\rho(\cdot)$ is the softmax function.
Image-grounded Text Generation loss trains the fusion module to generate texts, given input images as the condition~\cite{DBLP:conf/icml/0008LSH23,DBLP:journals/corr/abs-2305-06500}. For the image-text pairs in the pre-training dataset, each image $I$ corresponds to a text sentence $\mathbf{y}_{1:T} = \{y_1,..., y_T\}$ of length $T$. We employ a multimodal causal self-attention mask for multimodal encoder $f_\theta(\cdot)$ and multimodal adaptive fusion module $M_\theta(\cdot)$ to control the interaction between queries and text. The visual query functions as a prefix causal, ensuring that queries can attend to each other while excluding text tokens. Similarly, each text token $y$ can attend to all visual queries and preceding text tokens. The loss function is defined as:
\begin{equation}
\mathcal{L}_{itg}= - \sum_{t=1}^T \log M_\theta (f_\theta(\left(y_t \mid  y_{<t},I\right)).
\end{equation}

\subsection{Retrieval-Augmented Retriever Training}\label{3.4}

During the retrieval stage, the retriever utilizes the question $x_q$ to retrieve relevant knowledge from multimodal knowledge base $\mathcal{KB}$. To achieve this, we apply the multimodal encoder $f_\theta(\cdot)$, which encodes the question $x_q$ along with all latent multimodal knowledge into an embedding space to identify the $\operatorname{Top-K}$ most relevant candidates, illustrated in Fig.~\ref{The framework} retrieval stage. We use contrastive learning to construct positive and negative samples for training. The knowledge type consists primarily of three components: image $k^I$, text $k^T$, and image-text $k^{IT}$. Thus, the $l$-th example in the dataset is represented as $(x_l,y_l,\{\hat{k}^I_i, \hat{k}^T_i,\hat{k}^{IT}_i \}_l,\{\overline{k}^I_j,\overline{k}^T_j,\overline{k}^{IT}_j\}_l)$, where $\hat{k}_i$ is the $i$-th positive (image, text, image-text) sample and $\overline{k}_j$ represents $j$-th negative (image, text, image-text)  sample. For a batch of knowledge examples, we gather all associated positive and negative knowledge sources into a batch $\mathcal{K}_B = \{\{\hat{k}^I_i, \hat{k}^T_i,\hat{k}^{IT}_i \}_1, \{\overline{k}^I_j,\overline{k}^T_j,\overline{k}^{IT}_j\}_1, ... ,\{\overline{k}^I_j,\overline{k}^T_j,\overline{k}^{IT}_j\}_B \}$. The multimodal encoder is responsible for encoding the multimodal feature representations and aligning the questions and knowledge within the unified semantic space. This alignment facilitates identifying the proximity between a question and its corresponding knowledge through contrastive learning. The objective function is defined as follows: 
\begin{equation}
\mathcal{L}_{con}=-\operatorname{log} \frac{\operatorname{exp}(f_\theta(x_q) \cdot f_\theta(\hat{k}^I;\hat{k}^T;\hat{k}^{IT}))}{\sum\limits_{k \in \mathcal{K}_B} \operatorname{exp}(f_\theta(x_q) \cdot f_\theta(k^I;k^T;k^{IT}))},
  \label{eq2}
\end{equation}
where $f_\theta(\cdot)$ is the multimodal encoder and $\mathcal{K}_B$ is a batch of knowledge sources. We use the multimodal encoder trained to encode text, image, and image-text features, and apply Maximum Inner Product Search~(MIPS)\cite{DBLP:conf/icml/GuoSLGSCK20} to select $\operatorname{Top-K}$ from knowledge base $\mathcal{KB}$ as the relevant $\mathcal{K}_{ret}$, as shown in the following: 
\begin{equation}
  \operatorname{TopK} (\mathcal{K}_{ret} \mid x_q)  = \underset{k \in \mathcal{KB}}{\operatorname{TopK}} \{f_\theta(x_q)\cdot f_\theta(k^I;k^T;k^{IT}) \}.
    \label{eq3}
\end{equation}
Although the retriever is more efficient for many retrieval tasks, its accuracy is lower on open multimodal question answering. There are instances where certain knowledge is confusing and bears token-wise similarity to the question at the feature level, yet it fails to understand the question and cannot provide an answer in the semantic space~\cite{DBLP:conf/emnlp/ChenHCVC22}. For instance, consider the question \emph{``The Marina Bay Sands in Singapore is made of what building material?"}, and the relevant text knowledge \emph{``Singapore is also the new downtown of Singapore, built on reclaimed land."}. The question is about the building materials of a Singapore hotel, but this text is about the location of a Singapore hotel. Despite their token-wise similarity in the words, they cannot answer this question and cause confusion. To address this issue, we introduce a rank strategy to sort the $\operatorname{Top-K}$ candidates and exclude confusing samples, where $K$ is the maximum number of positive samples corresponding to the question. We select the $\operatorname{Top-K}$ samples, categorize them into positive and negative samples based on the ground truth, and then perform rank training. Since the Q-Former does not have a classifier, we input the multimodal features output $f_\theta(k^I;k^T;k^{IT})$ by the multimodal encoder into the fixed-parameter LLMs encoder and trainable classifier $\boldsymbol{z}$. The loss function is defined as:
\begin{equation}
\mathcal{L}_{cls}=CrossEntropy\left(\boldsymbol{z}(f_\theta(k^I,k^T,k^{IT})), \boldsymbol{y}\right),
\label{eq4}
\end{equation}
where $\boldsymbol{y}$ is the ground truth about the knowledge is relevant or not.

\subsection{Adaptive Selection Knowledge Generation}\label{3.5}
During the generation stage, the retrieved multimodal knowledge is combined with the question $x_q$ as an augmented input $[k_1,..., k_l, x_q]$, which is fed to the multimodal encoder and LLMs~\cite{DBLP:journals/corr/abs-2210-11416} to produce multimodal representation encoding and generate answers. We observe that existing methods~\cite{DBLP:conf/emnlp/ChenHCVC22,DBLP:conf/acl/0002FYHL23} directly rely on retrieval results without distinguishing the correctness of the retrieved knowledge, potentially leading to the utilization of incorrect, confusing, or irrelevant information. To address this, we propose an adaptive selection knowledge generation~(ASKG) strategy based on a question-and-answer formulation, shown in Fig.~\ref{The framework} generation stage. ASKG strategy enables the generator to go beyond mere word similarity between the question and the retrieved knowledge, allowing it to grasp the semantic information of the question and identify which piece of knowledge contains the answer. Specifically, we manually construct question-and-answer data to enable the model to discriminate the relevance of multimodal knowledge, thereby utilizing the implicit capabilities of LLMs for knowledge filtering. Based on the original dataset, we select relevant knowledge as positive examples and irrelevant knowledge as negative examples. We create an ASKG enhanced dataset and combine the knowledge according to templates, with the identifier of the positive examples serving as the answer. 
The template for the question $\widetilde{x}_q$ is :
\emph{``We would like to request your feedback on ranking the questions according to their relevance to the references below. Relevance refers to the degree to which the reference can answer the question. The input format is Question: [content], Reference [knowledge ID]: [content]. The output format is: Related content is [knowledge ID]."}, and the answer $\widetilde{\mathbf{y}}$ is in the form of \emph{``The most relevant reference is Reference [knowledge ID]."}. 

We refer to the above enhanced dataset of questions and answers as $\widetilde{x}_q$ and $\widetilde{\mathbf{y}} = \{\widetilde{y}_1,..., \widetilde{y}_M\}$, where $M$ is the text length. Given the dataset question $x_q$ and ground-turth answer of length $T$, $\mathbf{y}_{1:T} = \{y_1,..., y_T\}$, as well as the constructed $\widetilde{x}_q$ and $\widetilde{\mathbf{y}}$, the generator $g_\theta(\cdot)$ utilizes attention over question $x_q$ and relevant knowledge $\mathcal{K}_{ret}$ encoded by multimodal encoder $f_\theta(\cdot)$ to generate textual outputs token by token. The final generation loss is defined by:
\begin{equation}
\begin{aligned}
    \mathcal{L}_{\text {gen}}=\sum_{i=1}^T & - \log g_\theta \left(y_i \mid y_{1:i-1},f_\theta(x_q,\mathcal{K}_{ret}) \right) \\ 
     +  \alpha \sum_{i=1}^M & -\log g_\theta \left(\widetilde{y}_i \mid \widetilde{y}_{1:i-1}, f_\theta(\widetilde{x}_q) \right),
\end{aligned}
\label{eq5}
\end{equation}
where $\alpha$ is the hyperparameter which will be discussed in section \ref{Sensitivity}. To give a clear illustration of RA-BLIP, we summarize the training pipeline in Algorithm~\ref{alg1}.

\begin{algorithm}[!t]
\caption{The Training pipeline for RA-BLIP.}
\begin{algorithmic}
\STATE \textbf{Retrieval Training Stage}
\STATE \hspace{0.25cm} \textbf{Input}: question $\{x_{qi}\}_{i=1}^{N}$, knowledge base $\mathcal{KB}$
\STATE \hspace{0.25cm} \textbf{for} sampled mini-batch $x_q$ and $\mathcal{K}_B$ \textbf{do}  
\STATE \hspace{0.5cm}  Compute contrastive loss $L_{con}$ by Eq.~(4)
\STATE \hspace{0.25cm} \textbf{end for}
\STATE \hspace{0.25cm} \textbf{Return} retrieval model $\theta_{ret}$ 
\STATE \hspace{0.25cm} \textbf{Input}: question $\{x_{qi}\}_{i=1}^{N}$, ground truth $\boldsymbol{y}$, $\operatorname{Top-K}$
\STATE \hspace{1.25cm} retrieved knowledge from $\theta_{ret}$
\STATE \hspace{0.25cm} \textbf{for} sampled mini-batch $x_q$, and $\operatorname{Top} \mathcal{K}_B$ \textbf{do}  
\STATE \hspace{0.5cm} Calculate crossentropy loss $L_{cls}$ by Eq.~(6)
\STATE \hspace{0.25cm} \textbf{end for}
\STATE \hspace{0.25cm} \textbf{Return} ranking model $\theta_{ran}$
\STATE \textbf{Generation Training Stage}
\STATE \hspace{0.25cm} \textbf{Input}: question $\{x_{qi}\}_{i=1}^{N}$, answer $\mathbf{y}$, $\mathcal{K}_{ret}$ from $\theta_{ran}$,
\STATE \hspace{1.25cm} ASKG datasets $\widetilde{x}_q$ and $\widetilde{\mathbf{y}}$ 
\STATE \hspace{0.25cm} \textbf{for} sampled mini-batch $x_q$, $\mathcal{K}_{ret}$ and $\widetilde{x}_q$ \textbf{do}  
\STATE \hspace{0.5cm}  Compute generation loss $L_{gen}$ by Eq.~(7)
\STATE \hspace{0.25cm} \textbf{end for}
\STATE \hspace{0.25cm} \textbf{Return} generation model $\theta_{gen}$
\end{algorithmic}
\label{alg1}
\end{algorithm}

\section{Experiments}

\subsection{Experimental Settings}
\subsubsection{Datasets}
We evaluate our method on three QA datasets: WebQA~\cite{DBLP:conf/cvpr/ChangCNGSB22}, MultimodalQA~\cite{DBLP:conf/iclr/TalmorYCLWAIHB21}, and MMCoQA~\cite{DBLP:conf/acl/LiLN22}. The details of these datasets are showcased in Table~\ref{tab: datasets}.

\begin{itemize}
\item{\textbf{WebQA}~\cite{DBLP:conf/cvpr/ChangCNGSB22} is a large-scale dataset for multimodal and multihop QA where all questions are knowledge-seeking queries that require two or more knowledge sources. Evaluation metrics are retrieval F1 and QA for assessing answer generation quality, which is measured as both fluency~(QA-FL) and accuracy~(QA-ACC). We calculate fluency through BARTScore~\cite{DBLP:conf/nips/YuanNL21} and evaluate accuracy via F1 and recall. The fluency score and accuracy score are multiplied $FL*Acc$ to calculate the overall score}.

\item{\textbf{MultimodalQA}~\cite{DBLP:conf/iclr/TalmorYCLWAIHB21} is a collection of multihop QA pairs that necessitate the fusion of knowledge from text, tables, and images. This dataset requires retrieval and reasoning across text, image, and tabular data types. The performance of MultimodalQA is measured by F1 score at the word level and the Exact Match~(EM) of the answers}.

\item{\textbf{MMCoQA}~\cite{DBLP:conf/acl/LiLN22} is the first dataset constructed for multimodal conversational QA tasks and aims to answer users’ questions with multimodal knowledge sources via multi-turn conversations.
It comprises multiple supervised signals, including decontextualized questions, answers, and corresponding evidence}.
\end{itemize}

\begin{table}[!t]
     \centering
    \caption{Overall details of downstream datasets.}
    \begin{tabular}{lccc}
        \toprule
        Dataset & Train & Dev & Test \\
        \midrule
        WebQA~\cite{DBLP:conf/cvpr/ChangCNGSB22} & 34.2K & 5K & 7.5K \\
        MultimodalQA~\cite{DBLP:conf/iclr/TalmorYCLWAIHB21} & 23.8K & 2.4K & 3.6K \\
        MMCoQA~\cite{DBLP:conf/acl/LiLN22} & 4.6K & 0.6K & 0.6K \\
        \bottomrule
    \end{tabular}
    
    \label{tab: datasets}
\end{table}

\subsubsection{Compared Methods}

For WebQA, MultimodalQA, and MMCoQA, we make comparisons with different baseline methods. The model parameter quantity comparison is shown in Table~\ref{parameters}. The number of Solar parameters is not published, and both Solar and SKURG use other models to exploit multimodal information without accounting for the parameter counts of other models. We have frozen LLM and ViT, focusing solely on training Q-Former, which has fewer trainable parameters and bfloat16 encoding. In order to verify the scaling law~\cite{kaplan2020scaling}, we selected more powerful FlanT5xxl for experiment. To compare LLMs with other methods of similar parameter magnitude, we utilized T5-base and T5-large as benchmarks for a fair comparison. Since T5-base and T5-large are not aligned with the model through pre-training, they need to be fine-tuned during training.

\begin{itemize}
\item{\textbf{VLP} \cite{DBLP:conf/cvpr/ChangCNGSB22,DBLP:conf/aaai/ZhouPZHCG20} pre-trains its transformer-based encoder-decoder with both textual and visual information. They first retrieve  knowledge based on the question and feed it into the model to generate answers. In addition, VLP integrates VinVL~\cite{DBLP:journals/corr/abs-2101-00529} to improve performance.}

\item{\textbf{MuRAG}~\cite{DBLP:conf/emnlp/ChenHCVC22} encodes the question and selects Top-K nearest neighbors from multimodal memory.
They are then fed into the backbone encoder-decoder to generate textual outputs token by token. The backbone model uses
T5-base \cite{DBLP:journals/jmlr/RaffelSRLNMZLL20} and ViT-large \cite{DBLP:conf/iclr/DosovitskiyB0WZ21}, respectively. }

\item{\textbf{SKURG}~\cite{DBLP:journals/corr/abs-2212-08632} takes multimodal information sources as input and encodes them separately, then utilizes an entity-centered fusion encoder to align the sources of different modalities via the shared entities and structured knowledge.
The method adopts OFA-base \cite{DBLP:conf/icml/WangYMLBLMZZY22} and BART-base \cite{DBLP:conf/acl/LewisLGGMLSZ20}. Besides, it integrates ELMo-based NER~\cite{DBLP:conf/acl/PetersABP17} and OpenNRE \cite{DBLP:conf/emnlp/HanGYYLS19} for entity and relation extraction.}


\item{\textbf{Solar}~\cite{DBLP:conf/acl/0002FYHL23} first converts multimodal inputs into textual data and then utilizes a T5~\cite{DBLP:journals/jmlr/RaffelSRLNMZLL20} to generate answers through retrieval, ranking, and decoding.
It retrieves and ranks the information using BERT~\cite{DBLP:conf/naacl/DevlinCLT19}. Additionally, it adopts BLIP \cite{DBLP:conf/icml/0001LXH22} for image caption generation and VinVL~\cite{DBLP:journals/corr/abs-2101-00529} for image-attribute feature extraction.}
\end{itemize}

\begin{table}[!t]
    \centering
    \caption{Comparison of parameter quantity. Parameter quantities of other methods refer to~\cite{DBLP:journals/corr/abs-2212-08632,DBLP:conf/acl/0002FYHL23}.}
    \begin{tabular}{lcc}
        \toprule
        Model & \#Trainable Params & \#Total Params \\
        \midrule
        VLP+VinVL~\cite{DBLP:conf/cvpr/ChangCNGSB22} & 220M & 220M \\
        MuRAG~\cite{DBLP:conf/emnlp/ChenHCVC22} & 527M & 527M \\
        SKURG~\cite{DBLP:journals/corr/abs-2212-08632} & 447M & 447M  \\
        ImplicitDecomp~\cite{DBLP:conf/iclr/TalmorYCLWAIHB21} & 1310M & 1310M \\
        RA-BLIP~(T5-base) & 387M & 1398M \\
        RA-BLIP~(T5-large) & 902M & 1913M \\
        RA-BLIP~(FlanT5xl) & 109M & 4.1B \\
        RA-BLIP~(FlanT5xxl) & 109M & 12.1B  \\
        \bottomrule
    \end{tabular}
    \label{parameters}
\end{table}

\subsubsection{Implementation Details}


Our method includes multimodal fusion pre-training, retrieval, ranking, and generation.
For WebQA and MultimodalQA, we follow all steps as mentioned above. For MMCoQA, we manually include the positive clues in the retrieval results without performing subsequent ranking, following the process used in previous work~\cite{DBLP:conf/acl/0002FYHL23}. We adopt InstructBLIP~\cite{DBLP:journals/corr/abs-2305-06500} with frozen EVA-ViT-g/14~\cite{DBLP:eva_clip} as well as LLM~(FlanT5xl and FlanT5xxl)~\cite{DBLP:journals/corr/abs-2210-11416} for generation. In order to compare LLMs with fewer than 1 billion parameters, we replaced the LLM backbone with T5-large and T5-base. Due to the inconsistent dimensions between T5 and Q-Former, we added a linear layer to align the dimensions and did not perform pre-training.
We keep the image encoder and the LLMs frozen, tuning only the Q-Former and the multimodal fusion module during the pre-training and retrieval stage. We adopt the multimodal encoder and LLM encoder as feature encoder at ranking stage. At the generation stage, we froze the image encoder as well as LLMs and only trained the Q-Former with ASKG. We froze the image encoder and FlanT5 during all training processes, as well as used bfloat16 encoding to achieve RA-BLIP with the fewest trainable parameters.

For pre-training, we pre-train the multimodal adaptive fusion module on the SBU dataset~\cite{DBLP:conf/nips/OrdonezKB11}. Our approach is to fix the parameters of the image encoder and Q-Former, and solely fine-tune the parameters of the multimodal adaptive fusion module. We use the AdamW optimizer and adopt cosine learning rate of 1e-5, warmup of 1K steps, and batch size $64$ for $10$ epochs. 
For fine-tuning, we use the AdamW optimizer with $\beta_1=0.9$, $\beta_2=0.99$, and a weight decay of $0.01$ uniformly for three datasets. For WebQA retrieval, we use a cosine learning rate of 1e-5, warmup of 1K steps, and batch size $4$ for $10$ epochs. For WebQA ranking, we select the top $10$ samples to train the model with a cosine learning rate of 1e-5, warmup of 1K steps and batch size $40$ for $5$ epochs. For generation, we adopt cosine learning rate of 1e-6, warmup of 1K steps, and batch size $4$ for $10$ epochs. We set learning rate of 5e-5 for T5-large and T5-base.

For MultimodalQA~\cite{DBLP:conf/iclr/TalmorYCLWAIHB21} and MMCoQA~\cite{DBLP:conf/acl/LiLN22}, we tested the results on the dev set of MultimodalQA as well as the dev and test sets of the MMCoQA. For MultimodalQA retrieval, we use a cosine learning rate of 1e-5, warmup of 1K steps, and a batch size $4$ for $10$ epochs. For MultimodalQA generation, we adopt a cosine learning rate of 1e-6, warmup of 1K steps, and a batch size of $8$ for $10$ epochs. We set learning rate of 1e-5 for T5-large and 5e-5 for T5-base. For MMCoQA, we manually include the positive clues in the retrieval results without performing subsequent ranking, following the process used in previous work \cite{DBLP:conf/acl/0002FYHL23}. 
We set the learning rate as 1e-5 and batch size as 32 for 10 epochs at the retrieval stage. Then, we use a cosine learning rate of 1e-6, warmup of 1K steps, and a batch size of $16$ for $10$ epochs at the generation stage. We use the standard evaluation protocol for each dataset and report the same metrics, as well as the random seeds are fixed for reproducibility.


\subsection{Main Results}











\begin{table}[!t]
\centering
\renewcommand{\arraystretch}{1.05}
\caption{Results of WebQA official test-set. $*$ represents LLM is FlanT5xl, while $\dag$ represents LLM is FlanT5xxl. Bold and \underline{underline} denote the best and previous SOTA results.}
\begin{tabular}{lccccc}
\toprule
Model & Retr-F1$\uparrow$ & QA-FL$\uparrow$ & QA-Acc$\uparrow$ & QA$\uparrow$ \\
\midrule
VLP~\cite{DBLP:conf/aaai/ZhouPZHCG20} & 0.69 & 42.6 & 36.7 & 22.6 \\

VLP+VinVL~\cite{DBLP:journals/corr/abs-2101-00529} & 0.71 & 44.2 & 38.9 & 24.1  \\

MuRAG~\cite{DBLP:conf/emnlp/ChenHCVC22} & 0.75 & 55.7 & 54.6 & 36.1 \\

SKURG~\cite{DBLP:journals/corr/abs-2212-08632} & 0.88 & 55.4 & 57.1 & 37.7 \\

Solar~\cite{DBLP:conf/acl/0002FYHL23} & \textbf{0.89} & \underline{60.9} & \underline{58.9} & \underline{40.9} \\

InstructBLIP$*$~\cite{DBLP:journals/corr/abs-2305-06500} & - & 51.7 & 59.0 & 31.4\\

InstructBLIP$\dag$~\cite{DBLP:journals/corr/abs-2305-06500} & - & 53.4 & 62.5 & 35.0\\

\midrule
RA-BLIP~(T5-base) & - & 62.6 & 59.7 &  41.6 \\

RA-BLIP~(T5-large) & - & 62.9 & 60.9 & 42.5 \\

RA-BLIP$*$ & 0.83 & 65.1 & 65.3 & 45.8 \\

RA-BLIP$\dag$ & \textbf{0.89} & \textbf{65.5} & \textbf{68.7} & \textbf{48.5} \\
\bottomrule
\end{tabular}
\label{table3}
\end{table}

\textbf{Results on WebQA.} We show the WebQA results in Table~\ref{table3}. We can see that RA-BLIP surpasses all baselines in terms of both QA and retrieval F1 scores. RA-BLIP~(FlanT5xl) achieves $45.8\%$ accuracy, which is $+4.9\%$ higher than the state-of-the-art Solar~\cite{DBLP:conf/acl/0002FYHL23}. Especially the metric QA-Acc is $+6.4\%$ higher, proving the model's powerful generation ability. Besides, RA-BLIP~(FlanT5xxl) beats SOTA Solar by $7.6\%$ on overall QA accuracy, which shows that RA-BLIP complies with scaling law~\cite{kaplan2020scaling} and can improve the generation accuracy by using more advanced LLM.  
In order to prove that it is our RA-BLIP framework rather than the advanced MLLM backbone that improves the generative performance, we conducted experiments on InstructBLIP~\cite{DBLP:journals/corr/abs-2305-06500} on WebQA. We used RA-BLIP's optimal 0.89 search result for InstructBLIP generation and found that its accuracy was $5.9\%$ lower than Solar, which further proves the effectiveness of RA-BLIP framework and ASKG. RA-BLIP~(T5-base) and~(T5-large) are not pre-trained to align with Q-Former, but they achieve 41.6\% and 42.5\% accuracy respectively based on 0.89 search results, surpassing Solar and proving it is the RA-BLIP framework rather than LLM that improves performance.
Compared with Solar and SKURG which require additional model assistance, RA-BLIP does not use additional models, but it also achieves very good results in retrieval and is $14\%$ higher than MuRAG, which similarly does not use additional models.





\begin{table}[!t]
\renewcommand\arraystretch{1.05}
\centering
\caption{MultimodalQA dev-set results. $*$ represents LLM is FlanT5xl, while $\dag$ represents LLM is FlanT5xxl. Single-Modal and Mutli-Modal respectively indicate whether reasoning relies on Single-Modal or Mutli-Modal knowledge. Bold and \underline{underline} denote the best and SOTA results, respectively.}
\centering
\begin{tabular}{lccccccc}
\toprule
\multirow{2}{*}{Model} & \multicolumn{2}{c}{Single-Modal} & \multicolumn{2}{c}{Mutli-Modal} & \multicolumn{2}{c}{All} \\ \cmidrule(lr){2-3} \cmidrule(lr){4-5} \cmidrule(l){6-7}
& EM & F1  & EM & F1 & EM & F1 \\ 
\midrule
AutoRouting~\cite{DBLP:conf/iclr/TalmorYCLWAIHB21} & 51.7  & 58.5 & 34.2 & 40.2 & 44.7 & 51.1 \\

ImplicitDecomp~\cite{DBLP:conf/iclr/TalmorYCLWAIHB21} & 51.6 & 58.4 & 44.6 & 51.2 & 48.8 & 55.5 \\

SKURG~\cite{DBLP:journals/corr/abs-2212-08632} & 66.1 & 69.7 & 52.5 & 57.2 & \underline{59.8} & 64.0 \\

Solar~\cite{DBLP:conf/acl/0002FYHL23} & \underline{69.7} & \underline{74.8} & \underline{55.5} & \underline{65.4} & \underline{59.8} & \underline{66.1} \\
\midrule
RA-BLIP~(T5-base) & 65.4 & 71.6 & 59.7 & 65.7 & 63.1 & 69.3\\

RA-BLIP~(T5-large) & 65.2 & 71.9 & 62.6 & \textbf{68.4} & 64.1 & 70.5\\

RA-BLIP* & \textbf{70.1} & \textbf{77.6} & \textbf{59.3} & 65.5 & \textbf{65.8} & \textbf{72.7}\\
RA-BLIP$\dag$ & 69.9 & 76.4 &  59.1 & 65.6 & 65.6 & 72.1 \\
\bottomrule
\end{tabular}

\label{multimodalqa}
\end{table}

\textbf{Results on MultimodalQA.} We demonstrate MultimodalQA results in Table~\ref{multimodalqa}. MultimodalQA contains tables and has many multihop questions that require combining multimodal information. RA-BLIP also improved EM and F1 by $6.0\%$ and $6.6\%$, respectively, compared to state-of-the-art Solar, which demonstrates the generative ability of our method in incorporating multihop knowledge. In addition, RA-BLIP's accuracy is ahead of SOTA Solar in both single-modality and multi-modality, demonstrating our model can well combine multiple contextual semantic knowledge for cross-modal reasoning. Both RA-BLIP~(T5-large) and RA-BLIP~(T5-base) surpass Solar, indicating that the performance improvements are due to the RA-BLIP framework rather than the underlying LLM. Notably, the accuracy of more powerful FlanT5xxl is lower than that of FlanT5xl, probably because the powerful LLM is overfitted.

\begin{table}[!t]
\centering
\renewcommand\arraystretch{1.05}
\caption{MMCoQA test-dev-set results. $*$ represents LLM is FlanT5xl, while $\dag$ represents LLM is FlanT5xxl. Bold and \underline{underline} denote the best and SOTA results, respectively.}
\begin{tabular}{lccccc}
\toprule
\multirow{2}{*}{Model} & \multicolumn{2}{c}{Dev} & \multicolumn{2}{c}{Test} \\ \cmidrule(lr){2-3} \cmidrule(l){4-5} 
& EM & F1  & EM & F1 \\
\midrule
ORConvQA~\cite{DBLP:conf/sigir/Qu0CQCI20} & 1.0  & 3.0 & 1.0 & 1.9 \\

ManyModelQA~\cite{DBLP:conf/aaai/HannanJB20} & 0.7 & 2.3 & 1.0 & 1.8 \\

MAE~\cite{DBLP:conf/acl/LiLN22} & 21.5 & 30.2 & 24.9 & 32.3 \\

Solar~\cite{DBLP:conf/acl/0002FYHL23} & \underline{56.8} & \underline{62.5} & \underline{57.3} & \underline{64.6} \\
\midrule
RA-BLIP* & \textbf{59.2} & \textbf{67.1} & \textbf{61.0} & \textbf{67.8} \\
RA-BLIP$\dag$ & 58.7 & 66.7 & 59.5 & 66.7 \\
\bottomrule
\end{tabular}
\label{table5}
\end{table}

\textbf{Results on MMCoQA.} Our results on MMCoQA are shown in Table~\ref{table5}. Compared with WebQA and MultimodalQA, MMCoQA requires the model to correctly incorporate dialog history and develop deep multimodal understanding and reasoning capabilities across multiple conversations. RA-BLIP achieves a margin of $3.7\%$ enhancement over the best Solar for the EM score and $3.2\%$ for the F1 score in the test split. These results suggest the generalization ability and versatility of our model. Due to the small dataset, RA-BLIP~(FlanT5xxl) has resulted in overfitting, which prevents further performance improvement.





\begin{figure}[htbp]
  \centering
   \includegraphics[width=1\linewidth]{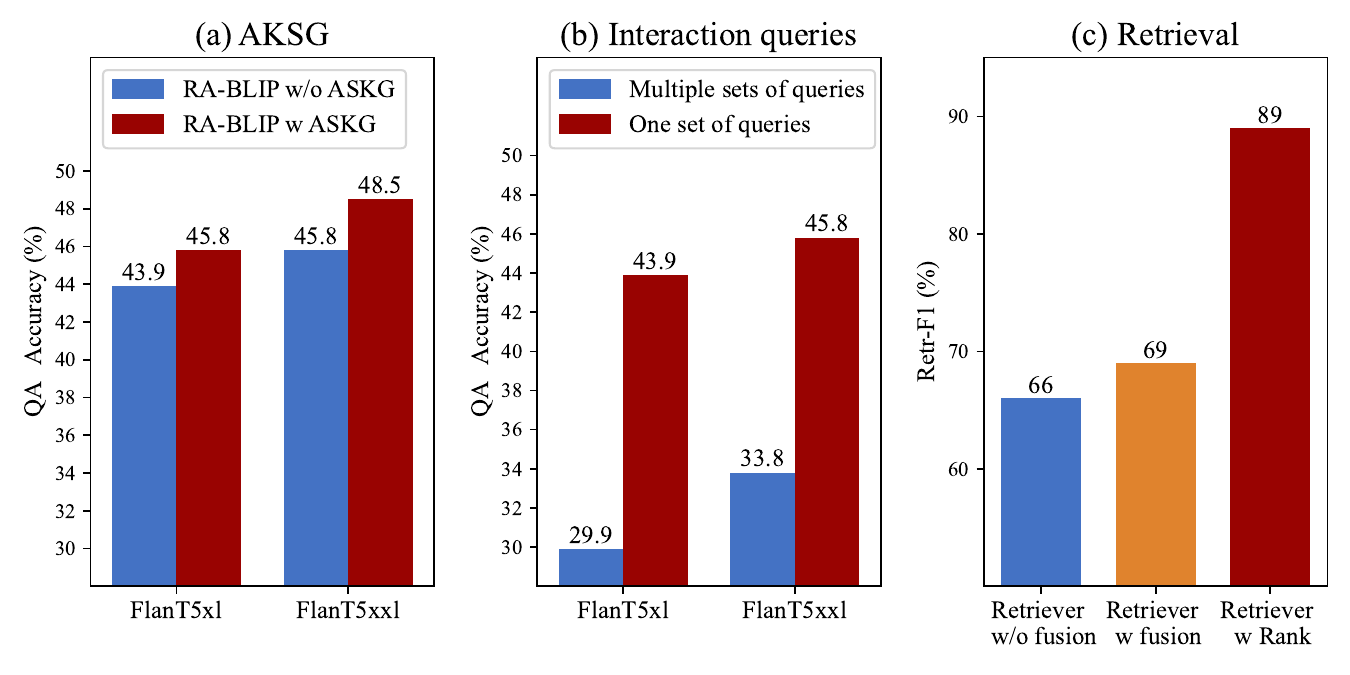}
   \caption{Ablation study for RA-BLIP generation and retrieval on WebQA.}
   \label{ablation}
\end{figure}



\subsection{Ablation Study}  

To analyze the effectiveness of our proposed method, we conducted comprehensive ablations on the WebQA dataset in both retrieval and generation stages. As shown in Fig.~\ref{ablation}~(a), RA-BLIP (FlanT5xl) and RA-BLIP (FlanT5xxl) with ASKG strategy improved by $1.9\%$ and $2.7\%$ respectively, which validates that employing ASKG strategy can assist LLMs in efficiently discerning the relevance of retrieved knowledge and effectively activate the implicit capabilities of more powerful LLMs. In Fig.~\ref{ablation}~(b), we compared the effects of one set of queries and multiple sets of queries, where one set of queries has a significant improvement. This proves that compared to multiple sets of queries that simply concatenate visual information from different images, one set of queries can better interact with and extract mixed visual information from multiple images at the feature level. We demonstrated the ablation results of RA-BLIP at the retrieval stage in Fig.~\ref{ablation}~(c). The result of the retriever without the multimodal adaptive fusion module exhibits lower performance than that of the complete retriever, which suggests that fusing vision and language in a unified semantic space is essential for multimodal retrieval. By utilizing a retrieval-rank strategy, the retrieval result is significantly improved $20\%$, demonstrating the necessity of denoising confusing knowledge through fine-level ranking. The retrieval-rank process employed in our method facilitates precise retrieval among candidate knowledge sources that are primarily relevant but potentially confusing.

\begin{figure}[!t]
  \centering
   \includegraphics[width=0.85\linewidth]{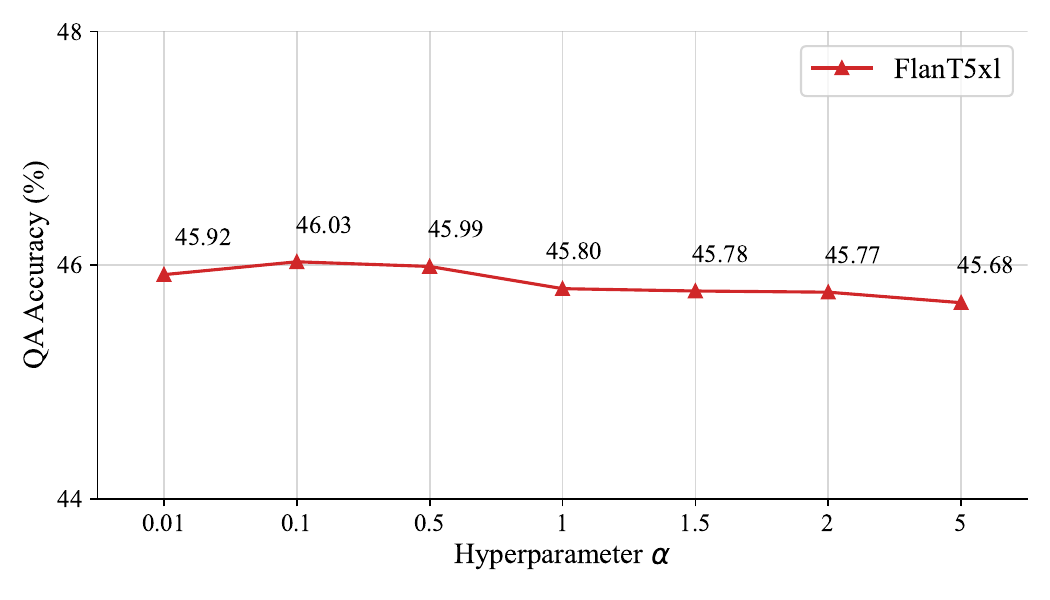}
   \caption{Influence of varying the hyperparameter $\alpha$ of generation loss.}
   \label{sensitivity}
\end{figure}

\subsection{Sensitivity Analysis}\label{Sensitivity}

The $\alpha$ is the trade-off hyperparameter of generation loss with ASKG strategy in Eq.~(7). We set the range of $\alpha$ from $0.01$ to $5$. According to Fig.~\ref{sensitivity}, we can see that even for such a large range, the difference between the best and the lowest results is less than $0.04\%$, indicating our method is robust and insensitive to this parameter.

\begin{figure}[!t]
   \centering
   \includegraphics[width=0.9\linewidth]{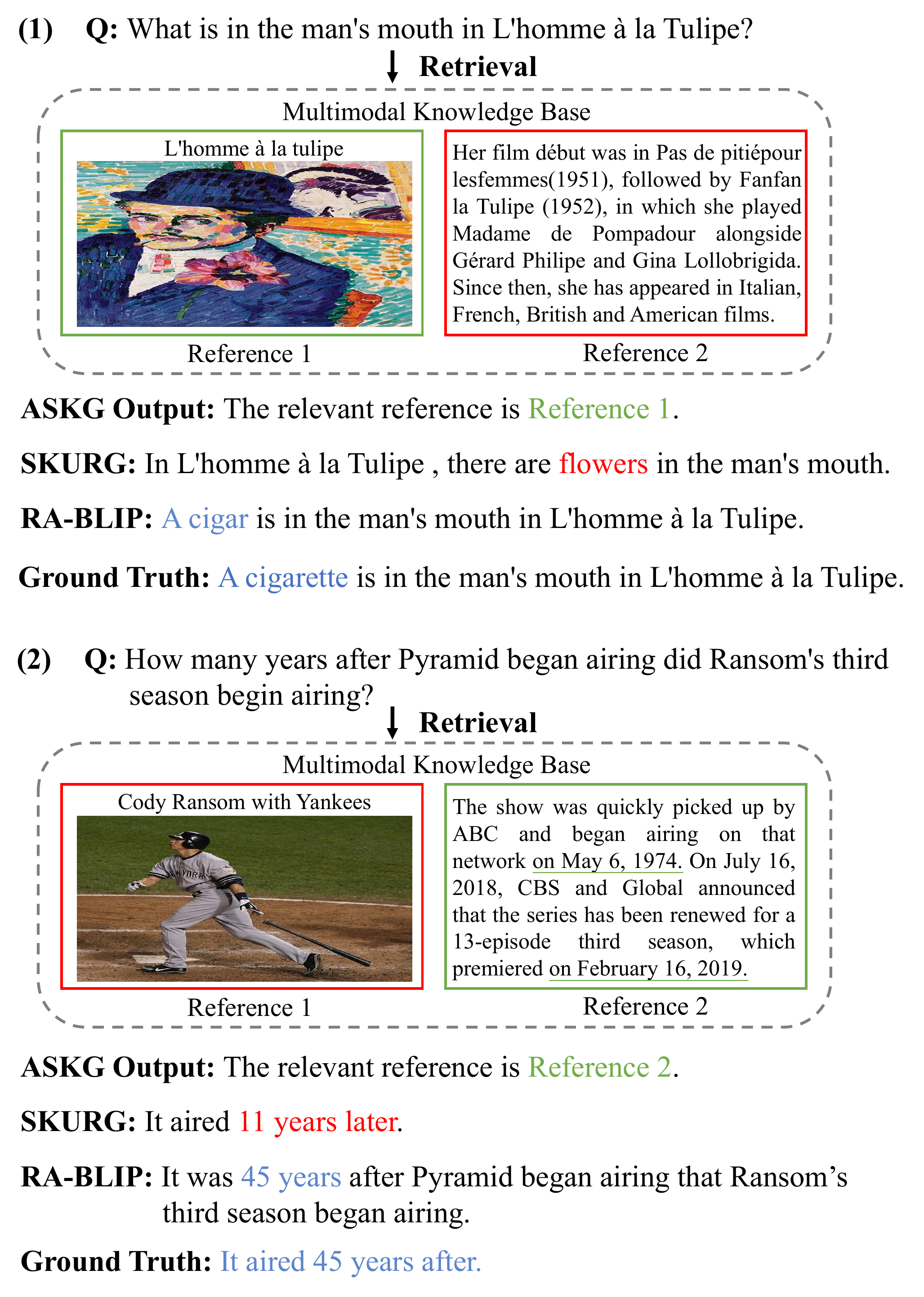}
   \caption{QA Examples. We demonstrate ASKG, RA-BLIP answers, SKURG~\cite{DBLP:journals/corr/abs-2212-08632} and ground truth. Relevant knowledge is in \textcolor[RGB]{114,173,67}{green window} and irrelevant is in \textcolor{red}{red window}. Relevant text in document is \textcolor[RGB]{114,173,67}{\underline{underlined}}.}
   \label{figure5}
\end{figure}

\subsection{Qualitative Results}

Fig.~\ref{figure5} illustrates four examples obtained by RA-BLIP and the baseline SKURG~\cite{DBLP:journals/corr/abs-2212-08632}. The retrieved relevant and confusing knowledge is listed, where the green boxes indicate the positive clue and the red boxes represent the negative cue. RA-BLIP outputs the correct answer, while SKURG generates the wrong answer under the identical conditions. This indicates RA-BLIP has more powerful abilities to comprehensively understand the retrieval information, regardless of textual or visual modality. Besides, through ASKG, our model autonomously judges the relevance of retrieved knowledge and selects relevant ones to generate more accurate answers.


\section{Conclusion}

In this paper, we propose a novel multimodal adaptive Retrieval-Augmented BLIP (RA-BLIP), a general retrieval-augmented framework for various classical MLLMs. RA-BLIP utilizes questions as instructions to extract visual features for less irrelevant interference. It incorporates a pre-trained multimodal adaptive fusion module to efficiently integrate information from both visual and textual modalities, thereby achieving question text-to-multimodal retrieval. Additionally, we introduce an adaptive selection knowledge generation strategy to make the generator autonomously discern the relevance of retrieved knowledge. Extensive experiments on multimodal multihop QA and ablation studies verify the effectiveness of RA-BLIP. In the future, we will explore image-multimodal retrieval and multimodal-multimodal retrieval to realize omnipotent retrieval-augmented models.


\bibliographystyle{IEEEtran}
\bibliography{RA_BLIP}

\end{document}